\documentclass[aps,prl,preprint,showpacs,showkeys,superscriptaddress]{revtex4-2}
\usepackage{amsmath,amssymb,amsfonts}
\usepackage{graphicx}
\usepackage{hyperref}
\usepackage{xcolor}
\usepackage{multirow}
\usepackage{booktabs}
\usepackage{caption}
\usepackage{pgfplots}
\usepackage{lipsum}
\usepackage{float}
\pgfplotsset{width=0.47\textwidth,height=0.33\textwidth,compat=1.9}

\begin{document}
	
	\title{Mass Spectrum of Heavy Mesons using WKB Approximation}
	\author{Bhaskar Jyoti Hazarika}
	\affiliation{On deputation, North Gauhati College, Guwahati, Assam, India}
	\affiliation{Center of Theoretical Studies, Pandu College, Guwahati 781012, India}
	\author{Tanmay Dev}
	\affiliation{Center of Theoretical Studies, Pandu College, Guwahati 781012, India}
	\affiliation{Department of Physics, Gauhati University, Guwahati 781014, India}

	\begin{abstract}
		In this study, we analyze the bound-state energy spectrum of quark-antiquark systems using the semiclassical WKB approximation. We consider the Cornell potential, which combines a linear confinement term with a Coulombic interaction, and perform a Taylor series expansion around an appropriately chosen point $r_{0}$ to simplify the potential near the classical turning points. This expansion enables an analytic treatment of the Schrödinger equation within the leading-order WKB framework. The resulting quantization condition yields approximate energy levels that capture the essential dynamics of quarkonium systems and illustrate the utility of the WKB method in modeling hadronic bound states.
	\end{abstract}
	\maketitle
	\section{Introduction}
	Mesons, as fundamental particles composed of a quark and an antiquark, are key to understanding the dynamics of the strong interaction described by quantum chromodynamics (QCD) \cite{sergeenko2019light}. As part of the hadron family, mesons provide a rich testing ground for theories of the strong force due to their relatively simple quark structure compared to baryons, which consist of three quarks. The study of meson mass spectra is crucial as it offers insights into the confinement mechanism in QCD, where quarks are bound together so tightly that they cannot be isolated. The mass spectra of mesons are determined by the interactions between the constituent quarks, mediated by gluons. These interactions are governed by the principles of QCD, a non-Abelian gauge theory that describes the strong force. Understanding meson masses helps in probing the non-perturbative regime of QCD, where the coupling constant becomes large, and perturbative methods fail. By studying mesons, we can test various potential models and approximation methods, leading to a better understanding of the strong interaction.
	\par Quarkonia are a special class of mesons composed of a quark and its own antiquark. The most well-known quarkonium states are charmonium($c\bar{c}$) and bottomonium ($b\bar{b}$). These states are particularly interesting because their relatively heavy quark content makes them more stable and allows for a clearer theoretical analysis. Charmonium and bottomonium states are bound by the same QCD interactions as lighter mesons, but their heavier mass results in different energy scales and more pronounced non-perturbative effects.
	\par
	Quarkonia states are important for several reasons:
	\begin{itemize}
		\item \textbf{Testing QCD Predictions:} The study of quarkonia allows for precise tests of QCD predictions in the non-perturbative regime.
		\item \textbf{Spectroscopy:} Quarkonium spectroscopy provides detailed information about the potential that binds quarks.
		\item \textbf{Decay Channels:} The decay channels of quarkonia are diverse and provide insights into weak and strong interaction processes.
	\end{itemize}
	\par Over the years, several models and computational techniques have been developed to study meson spectra. Early models, such as the quark model, provided a basic framework for understanding meson properties. However, more sophisticated models, like potential models and lattice QCD, have been developed to achieve better accuracy. The strong force, described by QCD, becomes weaker at short distances and stronger at long distances. This explains why quarks behave as quasi-free particles inside hadrons but cannot be isolated—an effect known as confinement. In systems with heavy quarks, like charm or bottom quarks, the motion is non-relativistic, making it possible to study them using potential models similar to those used for positronium. At short distances, the interquark potential resembles a Coulomb potential due to one-gluon exchange. At long distances, lattice QCD suggests a linearly rising potential due to confinement, though the full QCD potential is not exactly known. To model quarkonia (bound heavy quark–antiquark systems), a potential combining Coulomb, linear, and quadratic terms is used \cite{doi:10.1142/S0217732314501703}. This form not only reflects the key features of the strong interaction but also allows for near-analytical solutions useful in predicting quarkonium spectra and decay behaviors.The Cornell potential, a widely used potential model, effectively captures the behavior of the strong force at both short and long distances. This potential is given by:
	
	\[
	V(r) = -\frac{4\alpha_s}{3r} + B r
	\]
	
	where \(\alpha_s\) is the strong coupling constant, and $B$ is the QCD string tension. The Coulomb-like term dominates at short distances, reflecting asymptotic freedom, while the linear term dominates at long distances, representing confinement. Recent studies have applied the Wentzel-Kramers-Brillouin(WKB) approximation in conjunction with Pekeris type approximation scheme to solve such potentials and obtain meson mass spectra with good agreement to experimental data \cite{https://doi.org/10.1155/2020/5901464,bukor2023quarkonium}. Our work is also basically inspired by \citealt{https://doi.org/10.1155/2020/5901464} where we try a different approach in determining the masses of mesons. 
	
	\section{WKB approximation}
	\subsection{Introduction}
	The WKB approximation is a semi-classical method used to solve the Schrödinger equation for systems with smoothly varying potentials \cite{faustov2000algebraic, sergeenko2002zeroth, merzbacher1998quantum}. It is particularly useful for studying bound states where exact solutions are difficult to obtain. The method provides approximate expressions for the wave functions and energy levels, making it suitable for analyzing the mass spectra and decay constants of structures like quarkonia, tetraquarks, and pentaquarks etc. The Wentzel-Kramers-Brillouin (WKB) approximation is an effective method initially developed to find approximate solutions to the one-dimensional time-independent Schrödinger equation in the regime of large radial quantum numbers \cite{faustov2000algebraic, sergeenko2002zeroth, merzbacher1998quantum}. It is utilized to determine the finite wave functions and energy eigenvalues of relevant potentials. The WKB method has applications in studying quantum tunneling rates in a potential barrier, resonance behavior in a continuum, the exponential decay of unstable systems \cite{merzbacher1998quantum}, and quasi-normal nodes of black holes and quantum cosmology \cite{lu2018wkb}. A known limitation of the WKB approximation is its failure at the classical turning point, where momentum vanishes, a problem that can be mitigated using connection formulas \cite{griffiths2005introduction, merzbacher1998quantum}. The accuracy of the WKB method varies significantly for ground and low-lying states, depending on the potential function \cite{hruska1997accuracy}. Notably, the leading-order WKB approximation does not provide exact eigenvalues for the radial Schrödinger equation \cite{faustov2000algebraic}. This issue is addressed by replacing the orbital centrifugal barrier term $l(l+1)$ with $(l+1/2)^{2}$ in radial SE, known as the Langer correction \cite{langer1937connection}. This correction regularizes the WKB wave function at the origin and ensures correct asymptotic behavior for large radial quantum numbers \cite{sergeenko1996semiclassical}. In hadron spectroscopy, this method allows for a detailed analysis of the energy levels and wavefunctions of quarks within hadrons, providing critical insights into the strong interactions described by quantum chromodynamics (QCD).
	
	\par The Wentzel-Kramers-Brillouin (WKB) method is a powerful semiclassical approximation for solving the one-dimensional, time-independent Schrödinger equation:
	\begin{equation}
		-\frac{\hbar^2}{2m} \frac{d^2\psi(x)}{dx^2} + V(x)\psi(x) = E\psi(x),
	\end{equation}
	particularly effective in regions where the potential $V(x)$ varies slowly compared to the quantum wavelength. The method provides a bridge between classical and quantum mechanics and is valid in the limit $\hbar \to 0$ or, more practically, when the de Broglie wavelength $\lambda(x)$ varies slowly over a wavelength.
	
	\section*{2. Ansatz and Expansion}
	
	To apply the WKB method, we use the following ansatz for the wavefunction:
	\begin{equation}
		\psi(x) = \exp\left(\frac{i}{\hbar} \sigma(x)\right),
	\end{equation}
	where $\sigma(x)$ is a complex function encoding both phase and amplitude. Plugging this into the Schrödinger equation and simplifying using:
	\[
	\hbar^2 \frac{d^2}{dx^2} e^{\frac{i}{\hbar} \sigma(x)} = e^{\frac{i}{\hbar} \sigma(x)} \left[ i\hbar \sigma''(x) - (\sigma'(x))^2 \right],
	\]
	we arrive at the nonlinear equation:
	\begin{equation}
		-i\hbar \sigma''(x) + (\sigma'(x))^2 = p^2(x)
	\end{equation}
	where, $$p(x) = \sqrt{2m(E - V(x))}$$
	
	\section*{3. Leading Order (Semiclassical) Approximation}
	
	To solve the above equation, we expand $\sigma(x)$ in powers of $\hbar$:
	\begin{equation}
		\sigma(x) = \sigma_0(x) + \frac{\hbar}{i} \sigma_1(x) + \left(\frac{\hbar}{i}\right)^2 \sigma_2(x) + \cdots.
	\end{equation}
	Substituting into the above equation and equating powers of $\hbar$, the leading-order term (order $\hbar^0$) yields:
	\begin{equation}
		(\sigma_0'(x))^2 = p^2(x) \quad \Rightarrow \quad \sigma_0(x) = \pm \int p(x)\, dx.
	\end{equation}
	Thus, $\sigma_0(x)$ corresponds to the classical action. Retaining the next order (order $\hbar$), we find:
	\begin{equation}
		\sigma_1'(x) = -\frac{1}{2} \frac{p'(x)}{p(x)} \quad \Rightarrow \quad \sigma_1(x) = -\frac{1}{2} \ln p(x).
	\end{equation}

	Combining the leading and next-to-leading terms, we obtain the WKB solution in the classically allowed region ($E > V(x)$):
	\begin{equation}
		\begin{split}
		\psi(x) \approx \frac{C_1}{\sqrt{p(x)}} \exp\left( \frac{i}{\hbar} \int^x p(x')\, dx' \right) \\ + \frac{C_2}{\sqrt{p(x)}} \exp\left( -\frac{i}{\hbar} \int^x p(x')\, dx' \right).\\
		\end{split}
	\end{equation}
	
	In the classically forbidden region ($E < V(x)$), $p(x)$ becomes imaginary: $p(x) = i\kappa(x)$, where $\kappa(x) = \sqrt{2m(V(x) - E)}$, and the solution becomes:
	\begin{equation}
		\begin{split}
		\psi(x) \approx \frac{C_1'}{\sqrt{\kappa(x)}} \exp\left( -\frac{1}{\hbar} \int^x \kappa(x')\, dx' \right) \\ + \frac{C_2'}{\sqrt{\kappa(x)}} \exp\left( \frac{1}{\hbar} \int^x \kappa(x')\, dx' \right).\\
		\end{split}
	\end{equation}
	
	The WKB approximation is valid when the potential varies slowly on the scale of a wavelength:
	\begin{equation}
		\left| \frac{1}{p^2(x)} \frac{dp(x)}{dx} \right| \ll 1 \quad \Leftrightarrow \quad \left| \frac{d\lambda(x)}{dx} \right| \ll \lambda(x),
	\end{equation}
	where $\lambda(x) = h/p(x)$ is the de Broglie wavelength.

	In a 1D potential well with classical turning points at $x = a$ and $x = b$, continuity conditions at the turning points lead to the WKB quantization condition:
	\begin{equation}
		\int_b^a p(x)\, dx = \left(n + \frac{1}{2}\right) \pi \hbar.
	\end{equation}
	This yields energy eigenvalues approximately matching exact solutions in certain solvable potentials (e.g., the harmonic oscillator).

	The probability density in the semiclassical region is:
	\[
	|\psi(x)|^2 \propto \frac{1}{p(x)},
	\]
	which implies the particle spends more time (i.e., has higher probability) in regions where it moves slowly—consistent with classical mechanics.

	\section{Formalism}
	In this paper we have solved the Schrodinger equation for Cornel Potential using WKB approximation with the help of taylor series. Since both the quarks are heavy so we have done non-relativistic treatment. Advantages of using the WKB Approximation with Taylor Series Expansion are-
	\begin{itemize}
		\item \textbf{Improved Analytical Tractability}- The Taylor series expansion allows us to approximate complicated or non-solvable potentials (e.g., the Cornell or Killingbeck potential) near classical turning points, making the WKB integrals analytically solvable or simpler to evaluate.
		\item \textbf{Applicability to Complex Potentials}- Many quark-antiquark interaction potentials are non-trivial and lack closed-form solutions. By expanding the potential near turning points, the WKB method can handle potentials that are otherwise too complex for direct integration.
		\item \textbf{Better Accuracy for Low-Energy States}- The Taylor-expanded potential provides a more precise local description of the potential well, improving the accuracy of the WKB quantization condition even for lower quantum numbers (where WKB is typically less accurate).
		\item \textbf{Semi-Analytical Results}- Unlike purely numerical methods, this combined approach often yields semi-analytical expressions for energy levels, which are useful for parameter studies and physical interpretation.
		\item \textbf{Preservation of Physical Insight}- The WKB approximation maintains a clear connection between quantum and classical behavior, and using a Taylor series preserves this insight by showing how energy levels depend on the local curvature of the potential.
		\item \textbf{Reduction of Computational Effort}-
		Once formulated, this method is flexible and can be applied to other bound-state systems with similar potential forms, making it a reusable analytical tool.
	\end{itemize}
		
	The time-independent Schrödinger Equation with a reduced mass $\mu$ and wave-function $\psi(r,\theta,\phi)$ is given as 
	\begin{equation} \label{1}
		\begin{split}
			& \frac{\hbar^{2}}{2\mu}\left[\frac{1}{r^{2}}\left(r^{2}\frac{\partial}{\partial{r}}\right)+\frac{1}{r^{2} \sin{\theta} }\frac{\partial}{\partial{\theta}} \left(\sin{\theta}\frac{\partial}{\partial{\theta}}\right) \right. \\
			& \left. \qquad + \frac{1}{r^{2}\sin{\theta}^{2}} \frac{\partial^{2}}{\partial{\phi^{2}}}\right]\psi(r,\theta,\phi) \\
			&  +V(r) \psi(r,\theta, \phi) = E\psi(r,\theta,\psi) \\
		\end{split}
	\end{equation}
	The classical momenta is given by,
	\begin{equation} \label{2}
		P(r) = \left\{2\mu\left[E-V(r)-\frac{(l+1/2)^{2}\hbar^{2}}{2\mu r^{2}}\right]\right\}^{\frac{1}{2}}
	\end{equation}

	As meson is composite system of quark and  antiquark, therefore we have to take reduced mass of the system,
	$$\mu = \frac{m_{1}m_{2}}{m_{1}+m_{2}}$$
	
	The potential we use is the Cornell potential,
	\begin{equation}
		V(r)= -\frac{4\alpha_{s}}{3r}+br=-\frac{C}{r}+Br
	\end{equation}
	
	The standard WKB quantization condition \cite{hruska1997accuracy, sergeenko1996semiclassical} for two turning points ($r_{1},r_{2}$) is given as
	\begin{equation} \label{3}
		\int_{r_{1}}^{r_{2}}P(r)dr=\pi \hbar \left(n+\frac{1}{2}\right),\ r_{1}<r<r_{2} \quad n=0,1,2...
	\end{equation}
	
	\subsection{Solution of the Radial Schrödinger Equation Using Taylor Series expansion}
	In this section, we derive the solution for the bound states of the Schrödinger Equation by employing the WKB standard quantization condition, which involves substituting Eq.\eqref{2} Eq.\eqref{3}
	\begin{equation} \label{4}
		\begin{split}
			\int_{r_{1}}^{r_{2}} & \left\{2\mu \left[ E - B r + \frac{C}{r} - \frac{(l+1/2)^{2} \hbar^{2}}{2 \mu r^{2}} \right] \right\}^{1/2} dr \\
			& = \pi \hbar \left(n + \frac{1}{2}\right), \quad r_{1} < r < r_{2}, \quad n = 0,1,2..
		\end{split}
	\end{equation}
	As the solution of the Schrodinger equation gives bound state energy levels, the WKB approximation also provide the same through the quantum quantization condition given by Eq. \eqref{4}
	Equation\eqref{4} yields three turning points, denoted as $r_{1}$, $r_{2}$ and $r_{3}$, where $P(r)=0$. Here 1st two turning points gives real solution and 3rd one gives complex solution. Solving the multiple turning point problem posed by the potential presents mathematical hurdles. Firstly, determining the turning points of the polynomial equation in Eq.\eqref{4} necessitates algebraic techniques. Secondly, even without the centrifugal barrier term, evaluating the integral in Eq.\eqref{4} poses analytical challenges.
	
	We make a Taylor series expansion of the term $\frac{1}{r}$ and $\frac{1}{r^{2}}$ in Eq.\eqref{4} to get,
	\begin{equation} \label{6}
		\frac{1}{r} = \frac{1}{r_0} - \frac{1}{r_0^2}(r - r_0) + \frac{2}{r_0^3} \frac{(r - r_0)^2}{2!} - \frac{6}{r_0^4} \frac{(r - r_0)^3}{3!} + \cdots
	\end{equation}
	
	\begin{equation} \label{7}
			\frac{1}{r^2} = \frac{1}{r_0^2} - \frac{2}{r_0^3}(r - r_0) + \frac{6}{r_0^4} \frac{(r - r_0)^2}{2!} - \frac{24}{r_0^5} \frac{(r - r_0)^3}{3!} + \cdots
	\end{equation}

	The sensitivity of the approximation refers to how rapidly the Taylor series deviates from the actual function as we move away from the expansion point \( r_0 \).
	
	 The function \( \frac{1}{r} \) is relatively stable around \( r_0 \). The Taylor approximation remains close over a modest range.
	For \( \frac{1}{r^2} \), the derivatives grow more quickly, so sensitivity increases. The approximation deteriorates more rapidly.
	For \( \frac{1}{r^3} \), even small deviations from \( r_0 \) cause large differences between the approximation and actual value.
	For \( \frac{1}{r^n} \), the derivatives scale as:
	\[
	\frac{d^k}{dr^k} \left( \frac{1}{r^n} \right) \propto \frac{n(n+1)\cdots(n+k-1)}{r^{n+k}}
	\]
	This factorial growth in the numerator and increasing powers in the denominator make the approximation highly sensitive as \( n \) increases.

\begin{figure*}[h]
		\centering
		\includegraphics[width=1.0\textwidth]{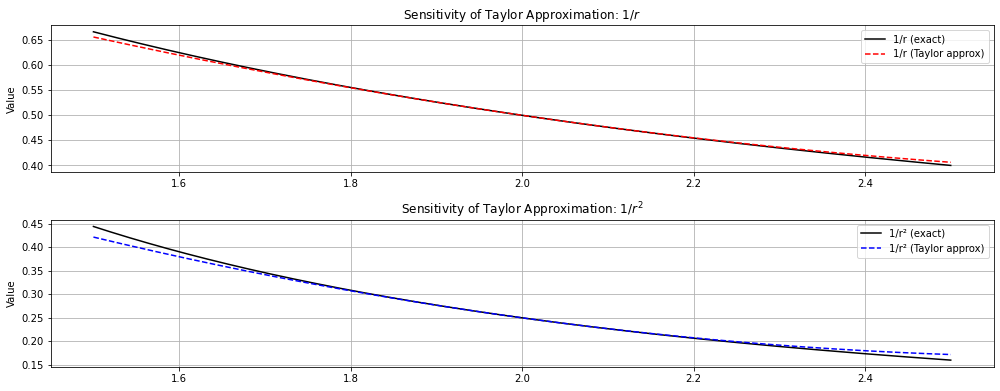}
		\caption{Comparison of exact functions (solid) and second-order Taylor approximations (dashed) for \( \frac{1}{r} \), and \( \frac{1}{r^2} \). Sensitivity increases with power.}
		\label{taylor}
	\end{figure*}

	The plots in \ref{taylor} illustrate the exact functions and their Taylor approximations around \( r_0 = 2 \) with a range of \( r \in [1.5, 2.5] \).

	The accuracy of Taylor approximations significantly depends on both the power of the function and the proximity to the expansion point \( r_0 \). As the power \( n \) increases in functions like \( 1/r^n \), the sensitivity to deviations from \( r_0 \) increases due to the rapid growth of higher-order derivatives. This has direct implications in physical modeling, especially near singularities or steep gradients.

	\par The point $r_0$ serves as an expansion center where the potential is locally approximated, allowing simplification of the integral expression. The value of $r_0$ is treated as a variational or phenomenological parameter, selected to optimize the agreement with known spectra or minimize error in the WKB approximation.

	\par Neglecting the higher order terms in \eqref{6} and \eqref{7}, we obtain
	
	\begin{equation}
		\begin{split}
			\int_{r_{1}}^{r_{2}}P(r)dr = \int_{r_{1}}^{r_{2}} \sqrt{2\mu} \Bigg[ & E + \frac{3C}{r_{0}} - \frac{3C}{r_{0}^{2}} + \frac{6F}{r_{0}^{2}} \\
			& - \left(B + \frac{3C}{r_{0}^{2}} + \frac{8F}{r_{0}^{3}}\right)r \\
			& + \left(\frac{C}{r_{0}^{3}} + \frac{3F}{r_{0}^{4}}\right)r^{2} \Bigg]^{1/2} \, dr
		\end{split}
	\end{equation}
		where, 
	\begin{equation}
		c=\frac{4 \alpha_{s}}{3}
	\end{equation}
	\begin{equation}
		F=\frac{(l+1/2)^{2}\hbar^{2}}{2\mu}
	\end{equation}
	which can be expressed as
	
	\begin{equation} \label{8}
		\int_{r_{1}}^{r_{2}}P(r)dr=\int_{r_{1}}^{r_{2}} \sqrt{2\mu} \left[ \lambda+\mu r+\nu r^{2}\right]^{1/2}dr
	\end{equation}
	
	where
	\begin{equation}\label{9}
		\lambda= E+  \frac{4\alpha_{s}}{ r_{0}}- \frac{4\alpha_{s}}{r_{0}^{2}}+  \frac{6F}{r_{0}^{2}}
	\end{equation}
	
	\begin{equation} \label{10}
		\mu=B+   \frac{4\alpha_{s}}{ r_{0}^{2}}+\frac{8F}{r_{0}^{3}}
	\end{equation}
	
	\begin{equation} \label{11}
		\nu=  \frac{4\alpha_{s}}{3 r_{0}^{3}}+\frac{3F}{r_{0}^{4}}
	\end{equation}
	Putting RHS of Eq \eqref{8} in the standard form as,
	\begin{equation} \label{12}
		\int_{x_{1}}^{x_{2}} \sqrt{2\mu}\sqrt{(x_{2}-x)(x-x_{1})}dx
	\end{equation}
	Here $x_{1}$ and $x_{2}$ represents the solution of the quadratic equation in the Eq.\eqref{8}.
	Solving these and applying quantization condition we get,
	\begin{equation} \label{13}
		\begin{aligned}
			E = &\frac{\left(B + \frac{3c}{r_0^2} + \frac{8F}{r_0^3}\right)^2}{4 \left(\frac{c}{r_0^3} + \frac{3F}{r_0^4}\right)} 
			+ \frac{2 \left(\frac{c}{r_0^3} + \frac{3F}{r_0^4}\right) h \left(n + \frac{1}{2}\right)}{\sqrt{2 \mu} \pi} \\
			& - \frac{3c}{r_0} + \frac{3c}{r_0^2} - \frac{6F}{r_0^2}
		\end{aligned}
	\end{equation}
	
	where, 
	\begin{equation}
		c=\frac{4 \alpha_{s}}{3}
	\end{equation}
	Therefore,
	\begin{equation} \label{14}
		M=m_{q}+m_{\bar{q}}+E_{nl}
	\end{equation}

	\begin{equation} \label{15}
		\begin{aligned}
			M = m_{q} + m_{\bar{q}} & + \frac{\left(B + \frac{3c}{r_0^2} + \frac{8F}{r_0^3}\right)^2}{4 \left(\frac{c}{r_0^3} + \frac{3F}{r_0^4}\right)} \\
			& + \frac{2 \left(\frac{c}{r_0^3} + \frac{3F}{r_0^4}\right) h \left(n + \frac{1}{2}\right)}{\sqrt{2 \mu}\pi} \\
			& - \frac{3c}{r_0} + \frac{3c}{r_0^2} - \frac{6F}{r_0^2}
		\end{aligned}
	\end{equation}
	
	This expression provides the energy eigenvalues in terms of the quantum number \(n\), the parameters of the potential \(\alpha_{s}\), \(B\), and the approximate position of the turning point \(r_{0}\).
	
	\subsection{Solution of Radial Equation Using Numerical Method}
	\par We have adopted a numerical method for the same and compared these results with the analytical method in this work and other theoretical and experimental findings.
	\par The given function is 
	\begin{equation}
		P(r) = \left\{2\mu\left[E-V(r)-\frac{(l+1/2)^{2}\hbar^{2}}{2\mu r^{2}}\right]\right\}^{\frac{1}{2}}
	\end{equation}
	\par To numerically integrate this, we evaluate the integral over some interval $[r_{min},r_{max}]$, where $P(r)$ is real and non-negative. So we solve the final integral using composite simpson's rule.
	\begin{equation}
		\begin{aligned}
			\int_{r_{\text{min}}}^{r_{\text{max}}} P(r) \, dr &\approx \frac{h}{3} \Big[ P(r_0) + 4 \sum_{i=1,3,5,\dots}^{n-1} P(r_i) \\
			&\quad + 2 \sum_{i=2,4,6,\dots}^{n-2} P(r_i) + P(r_n) \Big]
		\end{aligned}
	\end{equation}
	
	where, $$ h=\frac{r_{max}-r_{min}}{n}, r_{i}=r_{min}+i.h$$
	
	\par Now we define a target function $G(E)$ such that,
	\begin{equation}
		G(E)=\int_{r_{1}}^{r_{2}}P(r)dr-\pi\hbar(n+\frac{1}{2})
	\end{equation}
	We use 	Bisection method to solve the equation for $G(E)=0$
	\section{Results and Discussion}
\begin{figure*}[t]
		\centering
		\includegraphics[width=0.9\linewidth]{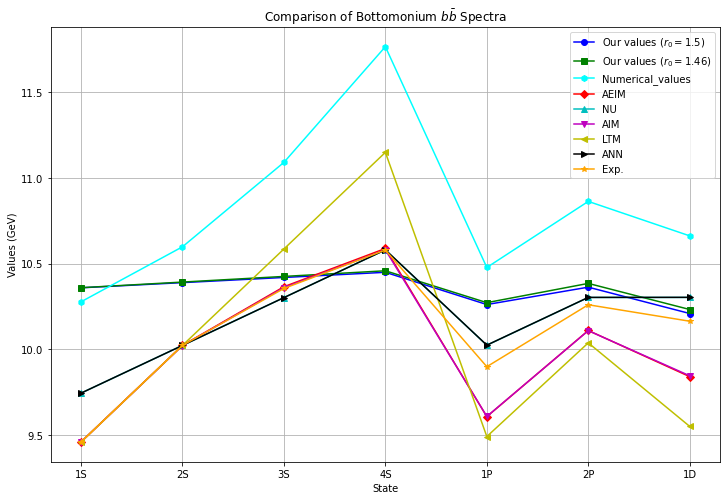}
		\caption{Bottomonium spectra: comparison with experimental data and other models presented}
		\label{fig:bottomonium}
	\end{figure*}

\begin{figure*}[t]
		\centering
		\includegraphics[width=0.9\linewidth]{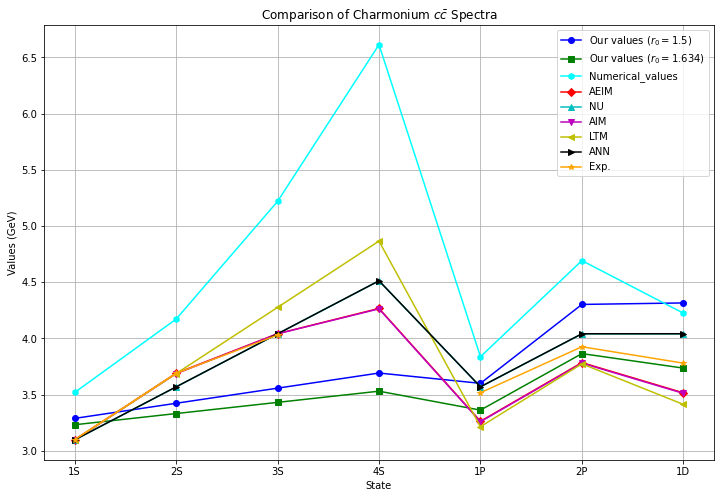}
		\caption{Charmonium spectra: comparison with experimental data and other available data}
		\label{fig:charmonium}
	\end{figure*}

\begin{figure*}[t]
		\centering
		\begin{minipage}{0.9\linewidth}
			\centering
			\includegraphics[width=\linewidth]{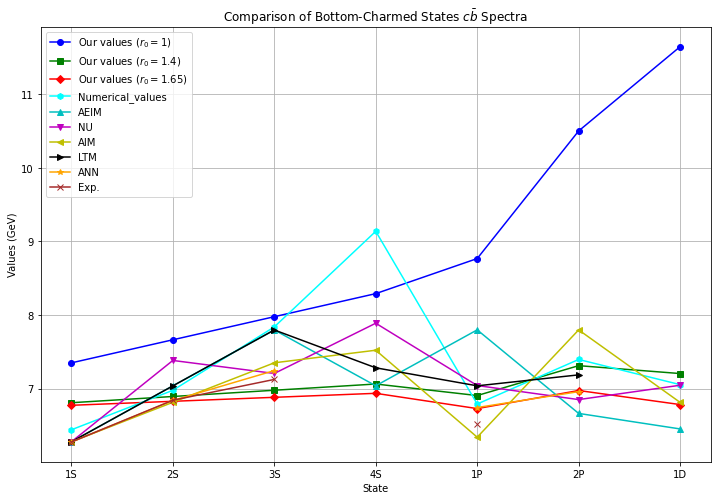}
			\caption{Bottom charmed spectra: comparison with experimental data and other available data}
			\label{fig:bottom_charmed}
		\end{minipage}
		
		\vspace{1em} 
		
		\begin{minipage}{0.9\linewidth}
			\centering
			\includegraphics[width=\linewidth]{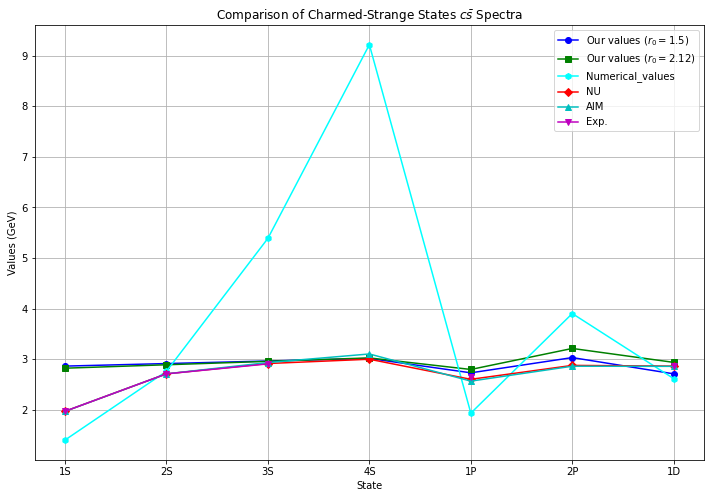}
			\caption{Charmed strange: comparison with experimental data and other available data}
			\label{fig:charmed_strange}
		\end{minipage}
	\end{figure*}

	The derived expression for the energy eigenvalues using the WKB approximation method showcases the dependence of the energy levels on the parameters of the potential and the quantum number. By comparing these theoretical predictions with previous findings like Nikoforov Uvarov(NU) method \cite{abu2019masses,ahmadov2021bound}, asymptotic iterative method(AIM) \cite{rani2018mass}, the Laplace transformation method (LTM) \cite{abu2018exact}, artificial neural network method(ANN) \cite{mutuk2019cornell}, and the analytical exact iterative method(AEIM) \cite{khokha2016quarkonium} in Table \ref{tab:comparison1}, \ref{tab:comparison2}, \ref{tab:comparison3} and \ref{tab:comparison4} we can validate the effectiveness of the WKB approximation for various mesonic systems. Omugbe et al. (2020) \cite{https://doi.org/10.1155/2020/5901464} applied the WKB approximation together with a Pekeris-type approximation to obtain meson mass spectra. We take $\alpha_s$ for charmonium and bottomium to be $0.278 $ and $0.185$ respectively \cite{doi:10.1142/1746}. We take input parameters same as our previous work \cite{hoque20202s,hazarika2011isgurwisefunctionqcdinspired}. For Bottomonium($b\bar{b}$) spectra our findings are in close proximity with the experimental \cite{patrignani2016review} and other models \cite{khokha2016quarkonium, abu2019masses, rani2018mass, abu2018exact, mutuk2019cornell} for $r_{0}=1.5$ and $r_{0}=1.46$. For Charmonium($c\bar{c}$) spectra it is evident from the Table \ref{tab:comparison2} our findings coincides with experimental \cite{patrignani2016review} and other models \cite{khokha2016quarkonium, abu2019masses, rani2018mass, abu2018exact, mutuk2019cornell} for $r_{0}=1.5$ and $r_{0}=1.634$. For Bottom-charmed($c\bar{b}$) states we calculated mass spectra for $r_{0}=1$, $r_{0}=1.45$ and $r_{0}=1.65$ where it coincides with experimental \cite{patrignani2016review} and other models \cite{khokha2016quarkonium, abu2019masses, rani2018mass, abu2018exact, mutuk2019cornell}. For Charmed-strange($c\bar{s}$) states we took $r_{0}=1.5$ and $r_{0}=2.12$ and our spectra shows similarity with available data. We have included figures(\ref{fig:bottomonium},\ref{fig:charmonium},\ref{fig:bottom_charmed},\ref{fig:charmed_strange}) to provide clearer visualization of the data and to effectively illustrate how the behavior of the quarkonium system differs from previous calculations and results.

	 In our analysis, some anomalies were observed, which likely arise from the high sensitivity of the Taylor series expansion to the choice of expansion point \( r_0 \). Even small deviations from this point can lead to significant errors, especially in functions involving inverse powers of \( r \). As the order of the function increases---such as in \( \frac{1}{r^2} \), \( \frac{1}{r^3} \), or more generally \( \frac{1}{r^n} \)---the corresponding derivatives grow rapidly, amplifying the sensitivity of the approximation.
	 
	 This behavior is rooted in the factorial scaling of the higher-order derivatives and the increasing power in the denominator, which together make the Taylor series increasingly unstable as \( n \) rises. Our graphical analysis around \( r_0 = 2 \) within the range \( r \in [1.5, 2.5] \) illustrates how the approximation deviates more strongly from the exact function with increasing \( n \).
	\section{Conclusion}

	In this work, as mesons are composite system of heavy quarks and antiquarks so we employed the non-relativistic Schrödinger equation with the Cornell potential to study the mass spectra of heavy quarkonia. By applying the WKB approximation along with a Taylor series expansion of the potential around a suitable point \( r_0 \), we derived semi-analytical expressions for the energy eigenvalues. We also obtained numerical solutions using the same potential model, which further supported the validity of our approach.The energy eigenvalue equation of the Schrödinger equation with the Cornell potential has been derived in a concise and elegant manner. This approximation scheme enables us to evaluate the WKB integral analytically for different angular momentum quantum number states.
	
	Our results show good agreement with experimental data and other theoretical methods such as the Nikiforov–Uvarov method, AIM, LTM, ANN, and AEIM, demonstrating the effectiveness of the WKB framework. However, the sensitivity of the Taylor expansion to the choice of \( r_0 \) plays a critical role—small deviations in \( r_0 \) can lead to noticeable discrepancies, especially for functions with higher inverse powers of \( r \). 
	
	This highlights the importance of careful parameter selection and motivates the inclusion of higher-order terms or alternative approximations in future work. Overall, our approach provides a reliable and semi-analytical framework for modeling meson mass spectra using potential-based methods.


	\section{Tables and Figures}

%
	
\begin{table*}[htbp]
	\captionsetup{width=\textwidth} 
	\centering
	\caption{Bottomonium $b\bar{b}$ spectra in $GeV (m_{b}=4.823GeV$, $\alpha_{s}=0.185$, $B=0.183GeV^{2}$)}
	\begin{tabular}{p{1.5cm} |p{1.5cm} p{1.5cm} p{1.5cm}| p{1.5cm} p{1.5cm} p{1.5cm} p{1.5cm} p{1.5cm} p{1.5cm}}
		\toprule
		State & $r_0=1.5$ & $r_0=1.46$ & Numerical values & AEIM [\cite{khokha2016quarkonium}] & NU [\cite{abu2019masses}] & AIM [\cite{rani2018mass}] & LTM [\cite{abu2018exact}] & ANN [\cite{mutuk2019cornell}] & Exp. [\cite{patrignani2016review}] \\
		\midrule
		1S & 10.3588 & 10.3593 & 10.2771 & 9.461 & 9.74473 & 9.46 & 9.46 & 9.745 & 9.46 \\
		2S & 10.3889 & 10.3922 & 10.5978 & 10.023 & 10.02315 & 10.022 & 10.023 & 10.023 & 10.023 \\
		3S & 10.419 & 10.4251 & 11.0897 & 10.365 & 10.30158 & 10.36 & 10.585 & 10.3016 & 10.355 \\
		4S & 10.4491 & 10.458 & 11.7639 & 10.588 & 10.58 & 10.58 & 11.148 & 10.58 & 10.579 \\
		1P & 10.2612 & 10.2722 & 10.4773 & 9.608 & 10.02406 & 9.609 & 9.492 & 10.0246 & 9.899 \\
		2P & 10.3625 & 10.3844 & 10.8623 & 10.11 & 10.30248 & 10.109 & 10.038 & 10.3029 & 10.26 \\
		1D & 10.209 & 10.2323 & 10.6614 & 9.841 & 10.30266 & 9.846 & 9.551 & 10.3032 & 10.164 \\
		\bottomrule
	\end{tabular}
	\label{tab:comparison1}
\end{table*}

\begin{table*}[htbp]
	 \captionsetup{width=\textwidth} 
	\centering
	\caption{Charmonium $c\bar{c}$ spectra in $GeV(m_{c}=1.209GeV$, $\alpha_{s}=0.278$, $B=0.183GeV^{2}$)}
	\begin{tabular}{p{1.5cm} |p{1.5cm} p{1.99cm} p{1.7cm}| p{1.5cm} p{1.5cm} p{1.5cm} p{1.5cm} p{1.5cm} p{1.5cm} p{1.2cm}}
		\toprule
		State & Our values ($r_0=1.5$) & Our values ($r_0=1.634$) & Numerical values & AEIM [\cite{khokha2016quarkonium}] & NU [\cite{abu2019masses}] & AIM [\cite{rani2018mass}] & LTM [\cite{abu2018exact}] & ANN [\cite{mutuk2019cornell}] & Exp. [\cite{patrignani2016review}] \\
		\midrule
		1S & 3.28746 & 3.23089 & 3.518 & 3.098 & 3.095481 & 3.095 & 3.096 & 3.0963 & 3.098 & 3.097 \\
		2S & 3.42199 & 3.33047 & 4.1705 & 3.689 & 3.567354 & 3.685 & 3.686 & 3.5681 & 3.688 & 3.686 \\
		3S & 3.55653 & 3.43004 & 5.2176 & 4.041 & 4.039226 & 4.04 & 4.275 & 4.04 & 4.029 & 4.039 \\
		4S & 3.69107 & 3.52962 & 6.6124 & 4.266 & 4.511098 & 4.262 & 4.865 & 4.5119 && \\
		1P & 3.59945 & 3.36117 & 3.8361 & 3.262 & 3.567735 & 3.258 & 3.214 & 3.5687 & 3.516 & 3.511 \\
		2P & 4.30156 & 3.86382 & 4.6913 & 3.784 & 4.039607 & 3.779 & 3.773 & 4.0406 & 3.925 & 3.927 \\
		1D & 4.31467 & 3.73495 & 4.2238 & 3.515 & 4.039683 & 3.51 & 3.412 & 4.0407 & 3.779 & 3.77 \\
		\bottomrule
	\end{tabular}
	\label{tab:comparison2}
\end{table*}

\begin{table*}[htbp]
	 \captionsetup{width=\textwidth} 
	\centering
	\caption{Bottom-charmed states $c\bar{b}$ spectra in $GeV(m_{c}=1.209GeV$, $m_{b}=4.823GeV$, $\alpha_{s}=0.223$, $B=0.183GeV^{2}$),}
		\begin{tabular}{p{1.5cm} |p{1.5cm} p{1.5cm} p{1.5cm} p{1.5cm}| p{1.5cm} p{1.5cm} p{1.5cm} p{1.5cm} p{1.5cm} p{1.5cm}}
			\toprule
			State & $r_0=1$ & $r_0=1.45$ & $r_0=1.65$ & Numerical values & AEIM [\cite{khokha2016quarkonium}] & NU [\cite{abu2019masses}] & AIM [\cite{rani2018mass}] & LTM [\cite{abu2018exact}] & ANN [\cite{mutuk2019cornell}] & Exp. Values [\cite{patrignani2016review}] \\
			\midrule
			1S & 7.35019 & 6.80828 & 6.77335 & 6.4405 & 6.277473 & 6.277 & 6.277 & 6.277 & 6.274 & 6.274 \\
			2S & 7.66394 & 6.89311 & 6.82762 & 6.977 & 7.037641 & 7.383 & 6.814 & 7.0372 & 6.839 & 6.842 \\
			3S & 7.97768 & 6.97795 & 6.88189 & 7.8387 & 7.797808 & 7.206 & 7.351 & 7.7973 & 7.245 & 7.125 \\
			4S & 8.29143 & 7.06278 & 6.93616 & 9.1385 & 7.038623 & 7.889 & 7.522 & 7.283 &  &  \\
			1P & 8.76811 & 6.90588 & 6.72846 & 6.7905 & 7.798791 & 7.042 & 6.34 & 7.0381 & 6.743 & 6.519 \\
			2P & 10.5028 & 7.31214 & 6.97443 & 7.3965 & 6.663 & 6.851 & 7.7983 & 7.187 & 6.959 &  \\
			1D & 11.6482 & 7.20501 & 6.78329 & 	7.0542 & 6.452 & 7.046 & 6.813 &  &  &  \\
			\bottomrule
		\end{tabular}%
	
	\label{tab:comparison3}
\end{table*}
	\begin{table*}[htbp]
	\captionsetup{width=\textwidth} 
	\centering
	\caption{Charmed-strange states $c\bar{s}$ spectra in $GeV(m_{c}=1.628GeV$, $m_{s}=0.419GeV $, $\alpha_{s}=0.223$, $B=0.183GeV^{2}$) }
	\begin{tabular}{p{1.5cm} |p{1.5cm} p{1.5cm} p{1.5cm}| p{1.5cm} p{1.5cm} p{3cm}}
		\toprule
		State & $r_0=1.5$ & $r_0=2.12$ & Numerical values & NU [\cite{abu2019masses}] & AIM [\cite{rani2018mass}] & Exp.[\cite{patrignani2016review}] \\
		\midrule
		1S & 2.862 & 2.82063 & 1.40067 & 1.969 & 1.968 & 1.96849 ± 0.00033 \\
		2S & 2.91241 & 2.88841 & 2.76007 & 2.709 & 2.709 & 2.709 \\
		3S & 2.96283 & 2.95619 & 5.38757 & 2.913 & 2.932 & 2.906 \\
		4S & 3.01324 & 3.02397 & 9.21527 & 2.998 & 3.102 &  \\
		1P & 2.72908 & 2.79562 & 1.93257 & 2.601 & 2.565 & 2.649 \\
		2P & 3.03036 & 3.21095 & 3.90097 & 2.876 & 2.86 &  \\
		1D & 2.70887 & 2.93566 & 2.61427 & 2.862 & 2.857 & 2.859 \\
		\bottomrule
	\end{tabular}
	
	\label{tab:comparison4}
\end{table*}

	\clearpage

	%


	\clearpage
	\def\bibsection{\section*{References}}

	\nocite{*}
	\bibliographystyle{apsrev4-2}
	\bibliography{reference2.bib}

\begin{thebibliography}{23}%
\makeatletter
\providecommand \@ifxundefined [1]{%
 \@ifx{#1\undefined}
}%
\providecommand \@ifnum [1]{%
 \ifnum #1\expandafter \@firstoftwo
 \else \expandafter \@secondoftwo
 \fi
}%
\providecommand \@ifx [1]{%
 \ifx #1\expandafter \@firstoftwo
 \else \expandafter \@secondoftwo
 \fi
}%
\providecommand \natexlab [1]{#1}%
\providecommand \enquote  [1]{``#1''}%
\providecommand \bibnamefont  [1]{#1}%
\providecommand \bibfnamefont [1]{#1}%
\providecommand \citenamefont [1]{#1}%
\providecommand \href@noop [0]{\@secondoftwo}%
\providecommand \href [0]{\begingroup \@sanitize@url \@href}%
\providecommand \@href[1]{\@@startlink{#1}\@@href}%
\providecommand \@@href[1]{\endgroup#1\@@endlink}%
\providecommand \@sanitize@url [0]{\catcode `\\12\catcode `\$12\catcode
  `\&12\catcode `\#12\catcode `\^12\catcode `\_12\catcode `\%12\relax}%
\providecommand \@@startlink[1]{}%
\providecommand \@@endlink[0]{}%
\providecommand \url  [0]{\begingroup\@sanitize@url \@url }%
\providecommand \@url [1]{\endgroup\@href {#1}{\urlprefix }}%
\providecommand \urlprefix  [0]{URL }%
\providecommand \Eprint [0]{\href }%
\providecommand \doibase [0]{https://doi.org/}%
\providecommand \selectlanguage [0]{\@gobble}%
\providecommand \bibinfo  [0]{\@secondoftwo}%
\providecommand \bibfield  [0]{\@secondoftwo}%
\providecommand \translation [1]{[#1]}%
\providecommand \BibitemOpen [0]{}%
\providecommand \bibitemStop [0]{}%
\providecommand \bibitemNoStop [0]{.\EOS\space}%
\providecommand \EOS [0]{\spacefactor3000\relax}%
\providecommand \BibitemShut  [1]{\csname bibitem#1\endcsname}%
\let\auto@bib@innerbib\@empty
\bibitem [{\citenamefont {Sergeenko}(2019)}]{sergeenko2019light}%
  \BibitemOpen
  \bibfield  {author} {\bibinfo {author} {\bibfnamefont {M.~N.}\ \bibnamefont
  {Sergeenko}},\ }\href@noop {} {\bibfield  {journal} {\bibinfo  {journal}
  {arXiv preprint arXiv:1909.10511}\ } (\bibinfo {year} {2019})}\BibitemShut
  {NoStop}%
\bibitem [{\citenamefont {Can\c{c}elik}\ and\ \citenamefont
  {G\"{o}n\"{u}l}(2014)}]{doi:10.1142/S0217732314501703}%
  \BibitemOpen
  \bibfield  {author} {\bibinfo {author} {\bibfnamefont {Y.}~\bibnamefont
  {Can\c{c}elik}}\ and\ \bibinfo {author} {\bibfnamefont {B.}~\bibnamefont
  {G\"{o}n\"{u}l}},\ }\href {https://doi.org/10.1142/S0217732314501703}
  {\bibfield  {journal} {\bibinfo  {journal} {Modern Physics Letters A}\
  }\textbf {\bibinfo {volume} {29}},\ \bibinfo {pages} {1450170} (\bibinfo
  {year} {2014})},\ \Eprint
  {https://arxiv.org/abs/https://doi.org/10.1142/S0217732314501703}
  {https://doi.org/10.1142/S0217732314501703} \BibitemShut {NoStop}%
\bibitem [{\citenamefont {Omugbe}\ \emph {et~al.}(2020)\citenamefont {Omugbe},
  \citenamefont {Osafile},\ and\ \citenamefont
  {Onyeaju}}]{https://doi.org/10.1155/2020/5901464}%
  \BibitemOpen
  \bibfield  {author} {\bibinfo {author} {\bibfnamefont {E.}~\bibnamefont
  {Omugbe}}, \bibinfo {author} {\bibfnamefont {O.~E.}\ \bibnamefont
  {Osafile}},\ and\ \bibinfo {author} {\bibfnamefont {M.~C.}\ \bibnamefont
  {Onyeaju}},\ }\href {https://doi.org/https://doi.org/10.1155/2020/5901464}
  {\bibfield  {journal} {\bibinfo  {journal} {Advances in High Energy Physics}\
  }\textbf {\bibinfo {volume} {2020}},\ \bibinfo {pages} {5901464} (\bibinfo
  {year} {2020})},\ \Eprint
  {https://arxiv.org/abs/https://onlinelibrary.wiley.com/doi/pdf/10.1155/2020/5901464}
  {https://onlinelibrary.wiley.com/doi/pdf/10.1155/2020/5901464} \BibitemShut
  {NoStop}%
\bibitem [{\citenamefont {Bukor}\ and\ \citenamefont
  {Tekel}(2023)}]{bukor2023quarkonium}%
  \BibitemOpen
  \bibfield  {author} {\bibinfo {author} {\bibfnamefont {B.}~\bibnamefont
  {Bukor}}\ and\ \bibinfo {author} {\bibfnamefont {J.}~\bibnamefont {Tekel}},\
  }\href@noop {} {\bibfield  {journal} {\bibinfo  {journal} {The European
  Physical Journal Plus}\ }\textbf {\bibinfo {volume} {138}},\ \bibinfo {pages}
  {499} (\bibinfo {year} {2023})}\BibitemShut {NoStop}%
\bibitem [{\citenamefont {Faustov}\ \emph {et~al.}(2000)\citenamefont
  {Faustov}, \citenamefont {Galkin}, \citenamefont {Tatarintsev},\ and\
  \citenamefont {Vshivtsev}}]{faustov2000algebraic}%
  \BibitemOpen
  \bibfield  {author} {\bibinfo {author} {\bibfnamefont {R.}~\bibnamefont
  {Faustov}}, \bibinfo {author} {\bibfnamefont {V.}~\bibnamefont {Galkin}},
  \bibinfo {author} {\bibfnamefont {A.}~\bibnamefont {Tatarintsev}},\ and\
  \bibinfo {author} {\bibfnamefont {A.}~\bibnamefont {Vshivtsev}},\ }\href@noop
  {} {\bibfield  {journal} {\bibinfo  {journal} {International Journal of
  Modern Physics A}\ }\textbf {\bibinfo {volume} {15}},\ \bibinfo {pages} {209}
  (\bibinfo {year} {2000})}\BibitemShut {NoStop}%
\bibitem [{\citenamefont {Sergeenko}(2002)}]{sergeenko2002zeroth}%
  \BibitemOpen
  \bibfield  {author} {\bibinfo {author} {\bibfnamefont {M.}~\bibnamefont
  {Sergeenko}},\ }\href@noop {} {\bibfield  {journal} {\bibinfo  {journal}
  {arXiv preprint quant-ph/0206179}\ } (\bibinfo {year} {2002})}\BibitemShut
  {NoStop}%
\bibitem [{\citenamefont {Merzbacher}(1998)}]{merzbacher1998quantum}%
  \BibitemOpen
  \bibfield  {author} {\bibinfo {author} {\bibfnamefont {E.}~\bibnamefont
  {Merzbacher}},\ }\href@noop {} {\emph {\bibinfo {title} {Quantum
  mechanics}}}\ (\bibinfo  {publisher} {John Wiley \& Sons},\ \bibinfo {year}
  {1998})\BibitemShut {NoStop}%
\bibitem [{\citenamefont {Lu}\ \emph {et~al.}(2018)\citenamefont {Lu},
  \citenamefont {Lv}, \citenamefont {Wang},\ and\ \citenamefont
  {Yang}}]{lu2018wkb}%
  \BibitemOpen
  \bibfield  {author} {\bibinfo {author} {\bibfnamefont {F.}~\bibnamefont
  {Lu}}, \bibinfo {author} {\bibfnamefont {B.}~\bibnamefont {Lv}}, \bibinfo
  {author} {\bibfnamefont {P.}~\bibnamefont {Wang}},\ and\ \bibinfo {author}
  {\bibfnamefont {H.}~\bibnamefont {Yang}},\ }\href@noop {} {\bibfield
  {journal} {\bibinfo  {journal} {Nuclear Physics B}\ }\textbf {\bibinfo
  {volume} {937}},\ \bibinfo {pages} {502} (\bibinfo {year}
  {2018})}\BibitemShut {NoStop}%
\bibitem [{\citenamefont {Griffiths}(2005)}]{griffiths2005introduction}%
  \BibitemOpen
  \bibfield  {author} {\bibinfo {author} {\bibfnamefont {D.}~\bibnamefont
  {Griffiths}},\ }\href {https://books.google.co.in/books?id=-BsvAQAAIAAJ}
  {\emph {\bibinfo {title} {Introduction to Quantum Mechanics}}},\ Pearson
  international edition\ (\bibinfo  {publisher} {Pearson Prentice Hall},\
  \bibinfo {year} {2005})\BibitemShut {NoStop}%
\bibitem [{\citenamefont {Hruska}\ \emph {et~al.}(1997)\citenamefont {Hruska},
  \citenamefont {Keung},\ and\ \citenamefont {Sukhatme}}]{hruska1997accuracy}%
  \BibitemOpen
  \bibfield  {author} {\bibinfo {author} {\bibfnamefont {M.}~\bibnamefont
  {Hruska}}, \bibinfo {author} {\bibfnamefont {W.-Y.}\ \bibnamefont {Keung}},\
  and\ \bibinfo {author} {\bibfnamefont {U.}~\bibnamefont {Sukhatme}},\
  }\href@noop {} {\bibfield  {journal} {\bibinfo  {journal} {Physical Review
  A}\ }\textbf {\bibinfo {volume} {55}},\ \bibinfo {pages} {3345} (\bibinfo
  {year} {1997})}\BibitemShut {NoStop}%
\bibitem [{\citenamefont {Langer}(1937)}]{langer1937connection}%
  \BibitemOpen
  \bibfield  {author} {\bibinfo {author} {\bibfnamefont {R.~E.}\ \bibnamefont
  {Langer}},\ }\href@noop {} {\bibfield  {journal} {\bibinfo  {journal}
  {Physical Review}\ }\textbf {\bibinfo {volume} {51}},\ \bibinfo {pages} {669}
  (\bibinfo {year} {1937})}\BibitemShut {NoStop}%
\bibitem [{\citenamefont {Sergeenko}(1996)}]{sergeenko1996semiclassical}%
  \BibitemOpen
  \bibfield  {author} {\bibinfo {author} {\bibfnamefont {M.}~\bibnamefont
  {Sergeenko}},\ }\href@noop {} {\bibfield  {journal} {\bibinfo  {journal}
  {Physical Review A}\ }\textbf {\bibinfo {volume} {53}},\ \bibinfo {pages}
  {3798} (\bibinfo {year} {1996})}\BibitemShut {NoStop}%
\bibitem [{\citenamefont {Abu-Shady}\ \emph {et~al.}(2019)\citenamefont
  {Abu-Shady}, \citenamefont {Abdel-Karim},\ and\ \citenamefont
  {Ezz-Alarab}}]{abu2019masses}%
  \BibitemOpen
  \bibfield  {author} {\bibinfo {author} {\bibfnamefont {M.}~\bibnamefont
  {Abu-Shady}}, \bibinfo {author} {\bibfnamefont {T.}~\bibnamefont
  {Abdel-Karim}},\ and\ \bibinfo {author} {\bibfnamefont {S.~Y.}\ \bibnamefont
  {Ezz-Alarab}},\ }\href@noop {} {\bibfield  {journal} {\bibinfo  {journal}
  {Journal of the Egyptian Mathematical Society}\ }\textbf {\bibinfo {volume}
  {27}},\ \bibinfo {pages} {14} (\bibinfo {year} {2019})}\BibitemShut {NoStop}%
\bibitem [{\citenamefont {Ahmadov}\ \emph {et~al.}(2021)\citenamefont
  {Ahmadov}, \citenamefont {Abasova},\ and\ \citenamefont
  {Orucova}}]{ahmadov2021bound}%
  \BibitemOpen
  \bibfield  {author} {\bibinfo {author} {\bibfnamefont {A.}~\bibnamefont
  {Ahmadov}}, \bibinfo {author} {\bibfnamefont {K.}~\bibnamefont {Abasova}},\
  and\ \bibinfo {author} {\bibfnamefont {M.~S.}\ \bibnamefont {Orucova}},\
  }\href@noop {} {\bibfield  {journal} {\bibinfo  {journal} {Advances in High
  Energy Physics}\ }\textbf {\bibinfo {volume} {2021}},\ \bibinfo {pages}
  {1861946} (\bibinfo {year} {2021})}\BibitemShut {NoStop}%
\bibitem [{\citenamefont {Rani}\ \emph {et~al.}(2018)\citenamefont {Rani},
  \citenamefont {Bhardwaj},\ and\ \citenamefont {Chand}}]{rani2018mass}%
  \BibitemOpen
  \bibfield  {author} {\bibinfo {author} {\bibfnamefont {R.}~\bibnamefont
  {Rani}}, \bibinfo {author} {\bibfnamefont {S.}~\bibnamefont {Bhardwaj}},\
  and\ \bibinfo {author} {\bibfnamefont {F.}~\bibnamefont {Chand}},\
  }\href@noop {} {\bibfield  {journal} {\bibinfo  {journal} {Communications in
  Theoretical Physics}\ }\textbf {\bibinfo {volume} {70}},\ \bibinfo {pages}
  {179} (\bibinfo {year} {2018})}\BibitemShut {NoStop}%
\bibitem [{\citenamefont {Abu-Shady}\ \emph {et~al.}(2018)\citenamefont
  {Abu-Shady}, \citenamefont {Abdel-Karim},\ and\ \citenamefont
  {Khokha}}]{abu2018exact}%
  \BibitemOpen
  \bibfield  {author} {\bibinfo {author} {\bibfnamefont {M.}~\bibnamefont
  {Abu-Shady}}, \bibinfo {author} {\bibfnamefont {T.}~\bibnamefont
  {Abdel-Karim}},\ and\ \bibinfo {author} {\bibfnamefont {E.}~\bibnamefont
  {Khokha}},\ }\href@noop {} {\bibfield  {journal} {\bibinfo  {journal} {arXiv
  preprint arXiv:1802.02092}\ } (\bibinfo {year} {2018})}\BibitemShut {NoStop}%
\bibitem [{\citenamefont {Mutuk}(2019)}]{mutuk2019cornell}%
  \BibitemOpen
  \bibfield  {author} {\bibinfo {author} {\bibfnamefont {H.}~\bibnamefont
  {Mutuk}},\ }\href@noop {} {\bibfield  {journal} {\bibinfo  {journal}
  {Advances in High Energy Physics}\ }\textbf {\bibinfo {volume} {2019}},\
  \bibinfo {pages} {3105373} (\bibinfo {year} {2019})}\BibitemShut {NoStop}%
\bibitem [{\citenamefont {Khokha}\ \emph {et~al.}(2016)\citenamefont {Khokha},
  \citenamefont {Abu-Shady},\ and\ \citenamefont
  {Abdel-Karim}}]{khokha2016quarkonium}%
  \BibitemOpen
  \bibfield  {author} {\bibinfo {author} {\bibfnamefont {E.}~\bibnamefont
  {Khokha}}, \bibinfo {author} {\bibfnamefont {M.}~\bibnamefont {Abu-Shady}},\
  and\ \bibinfo {author} {\bibfnamefont {T.}~\bibnamefont {Abdel-Karim}},\
  }\href@noop {} {\bibfield  {journal} {\bibinfo  {journal} {arXiv preprint
  arXiv:1612.08206}\ } (\bibinfo {year} {2016})}\BibitemShut {NoStop}%
\bibitem [{\citenamefont {Fayyazuddin}\ and\ \citenamefont
  {Riazuddin}(1992)}]{doi:10.1142/1746}%
  \BibitemOpen
  \bibfield  {author} {\bibinfo {author} {\bibnamefont {Fayyazuddin}}\ and\
  \bibinfo {author} {\bibnamefont {Riazuddin}},\ }\href
  {https://doi.org/10.1142/1746} {\emph {\bibinfo {title} {A Modern
  Introduction to Particle Physics}}}\ (\bibinfo  {publisher} {WORLD
  SCIENTIFIC},\ \bibinfo {year} {1992})\ \Eprint
  {https://arxiv.org/abs/https://worldscientific.com/doi/pdf/10.1142/1746}
  {https://worldscientific.com/doi/pdf/10.1142/1746} \BibitemShut {NoStop}%
\bibitem [{\citenamefont {Hoque}\ \emph {et~al.}(2020)\citenamefont {Hoque},
  \citenamefont {Hazarika},\ and\ \citenamefont {Choudhury}}]{hoque20202s}%
  \BibitemOpen
  \bibfield  {author} {\bibinfo {author} {\bibfnamefont {R.}~\bibnamefont
  {Hoque}}, \bibinfo {author} {\bibfnamefont {B.}~\bibnamefont {Hazarika}},\
  and\ \bibinfo {author} {\bibfnamefont {D.}~\bibnamefont {Choudhury}},\
  }\href@noop {} {\bibfield  {journal} {\bibinfo  {journal} {The European
  Physical Journal C}\ }\textbf {\bibinfo {volume} {80}},\ \bibinfo {pages}
  {1213} (\bibinfo {year} {2020})}\BibitemShut {NoStop}%
\bibitem [{\citenamefont {Hazarika}\ and\ \citenamefont
  {Choudhury}(2011)}]{hazarika2011isgurwisefunctionqcdinspired}%
  \BibitemOpen
  \bibfield  {author} {\bibinfo {author} {\bibfnamefont {B.~J.}\ \bibnamefont
  {Hazarika}}\ and\ \bibinfo {author} {\bibfnamefont {D.~K.}\ \bibnamefont
  {Choudhury}},\ }\href {https://arxiv.org/abs/1112.2800} {\bibinfo {title}
  {Isgur-wise function in a qcd inspired potential model with wkb
  approximation}} (\bibinfo {year} {2011}),\ \Eprint
  {https://arxiv.org/abs/1112.2800} {arXiv:1112.2800 [hep-ph]} \BibitemShut
  {NoStop}%
\bibitem [{\citenamefont {Patrignani}(2016)}]{patrignani2016review}%
  \BibitemOpen
  \bibfield  {author} {\bibinfo {author} {\bibfnamefont {C.}~\bibnamefont
  {Patrignani}},\ }\href@noop {} {\bibfield  {journal} {\bibinfo  {journal}
  {Chinese Physics C, High Energy Physics and Nuclear Physics}\ }\textbf
  {\bibinfo {volume} {40}} (\bibinfo {year} {2016})}\BibitemShut {NoStop}%
\bibitem [{\citenamefont {Mutuk}(2018)}]{https://doi.org/10.1155/2018/8095653}%
  \BibitemOpen
  \bibfield  {author} {\bibinfo {author} {\bibfnamefont {H.}~\bibnamefont
  {Mutuk}},\ }\href {https://doi.org/https://doi.org/10.1155/2018/8095653}
  {\bibfield  {journal} {\bibinfo  {journal} {Advances in High Energy Physics}\
  }\textbf {\bibinfo {volume} {2018}},\ \bibinfo {pages} {8095653} (\bibinfo
  {year} {2018})},\ \Eprint
  {https://arxiv.org/abs/https://onlinelibrary.wiley.com/doi/pdf/10.1155/2018/8095653}
  {https://onlinelibrary.wiley.com/doi/pdf/10.1155/2018/8095653} \BibitemShut
  {NoStop}%
\end{thebibliography}%
	
\end{document}